\documentclass[acmsmall]{acmart}
\usepackage{comment}
\usepackage{tabularx}
\usepackage{booktabs} 
\usepackage{caption} 
\usepackage{multirow}
\usepackage{multicol}
\usepackage[utf8]{inputenc}

\AtBeginDocument{%
  \providecommand\BibTeX{{%
    \normalfont B\kern-0.5em{\scshape i\kern-0.25em b}\kern-0.8em\TeX}}}

\setcopyright{acmcopyright}
\acmDOI{XXXXXXX.XXXXXXX}

\begin{document}

\title{Security and Privacy of Digital Twins for Advanced Manufacturing: A Survey}

\author{Alexander D. Zemskov}
\email{adz36@case.edu}
\affiliation{%
  \institution{Case Western Reserve University}
  \city{Cleveland}
  \state{Ohio}
  \country{USA}
}
\author{Yao Fu}
\email{yxf484@case.edu}
\affiliation{%
  \institution{Case Western Reserve University}
  \city{Cleveland}
  \state{Ohio}
  \country{USA}
}
\author{Runchao Li}
\email{rxl685@case.edu}
\affiliation{%
  \institution{Case Western Reserve University}
  \city{Cleveland}
  \state{Ohio}
  \country{USA}
}

\author{Xufei Wang}
\email{xxw512@case.edu}
\affiliation{%
  \institution{Case Western Reserve University}
  \city{Cleveland}
  \state{Ohio}
  \country{USA}
}

\author{Vispi Karkaria}
\email{vispikarkaria2026@u.northwestern.edu}
\affiliation{%
  \institution{Northwestern University}
  \city{Evanston}
  \state{Illinois}
  \country{USA}
}

\author{Ying-Kuan Tsai}
\email{yingkuan.tsai@northwestern.edu}
\affiliation{%
  \institution{Northwestern University}
  \city{Evanston}
  \state{Illinois}
  \country{USA}
}

\author{Wei Chen}
\email{weichen@northwestern.edu}
\affiliation{%
  \institution{Northwestern University}
  \city{Evanston}
  \state{Illinois}
  \country{USA}
}

\author{Jianjing Zhang}
\email{jxz170heh@case.edu}
\affiliation{%
  \institution{Case Western Reserve University}
  \city{Cleveland}
  \state{Ohio}
  \country{USA}
}

\author{Robert Gao}
\email{robert.gao@case.edu}
\affiliation{%
  \institution{Case Western Reserve University}
  \city{Cleveland}
  \state{Ohio}
  \country{USA}
}

\author{Jian Cao}
\email{jcao@northwestern.edu}
\affiliation{%
  \institution{Northwestern University}
  \city{Evanston}
  \state{Illinois}
  \country{USA}
}
\author{Kenneth A. Loparo}
\email{kal4@case.edu}
\affiliation{%
  \institution{Case Western Reserve University}
  \city{Cleveland}
  \state{Ohio}
  \country{USA}
}
\author{Pan Li}
\email{lipan@ieee.org}
\affiliation{%
  \institution{Case Western Reserve University}
  \city{Cleveland}
  \state{Ohio}
  \country{USA}
}

\renewcommand{\shortauthors}{Zemskov et al.}

\begin{abstract}
  In Industry 4.0, the digital twin is one of the emerging technologies, offering simulation abilities to predict, refine, and interpret conditions and operations, where it is crucial to emphasize a heightened concentration on the associated security and privacy risks. To be more specific, the adoption of digital twins in the manufacturing industry relies on integrating technologies like cyber-physical systems, the Industrial Internet of Things, virtualization, and advanced manufacturing.  The interactions of these technologies give rise to numerous security and privacy vulnerabilities that remain inadequately explored.  Towards that end, this paper analyzes the cybersecurity threats of digital twins for advanced manufacturing in the context of data collection, data sharing, machine learning and deep learning, and system-level security and privacy.  We also provide several solutions to the threats in those four categories that can help establish more trust in digital twins.
\end{abstract}


\ccsdesc[500]{Advanced Manufacturing~Digital Twin}
\ccsdesc[300]{Advanced Manufacturing~Security}
\ccsdesc[300]{Advanced Manufacturing~Privacy}

\keywords{data collection, data sharing, machine learning, deep learning, system-level security and privacy}


\maketitle

\section{Introduction}

\begin{figure}
\centering
\includegraphics[width=0.5\textwidth]{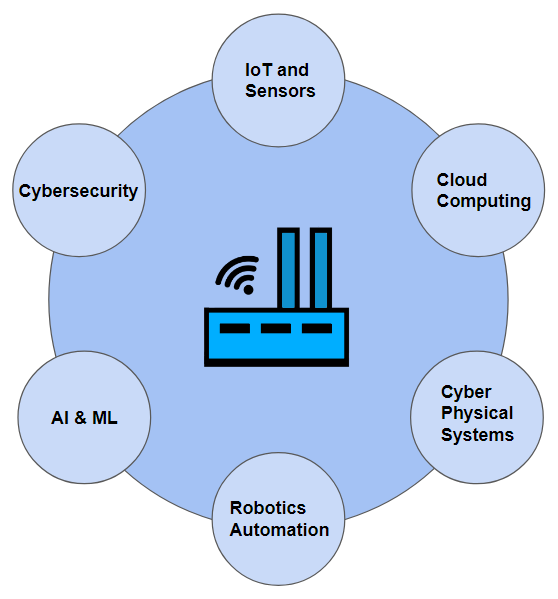}
\caption{Key Components of Advanced Manufacturing}
\label{fig:smartmanufacturing}
\end{figure}

In recent years, the global landscape of manufacturing has undergone a significant transformation, spurred by the integration of cutting-edge technologies and data-driven solutions. This paradigm shift, often referred to as advanced manufacturing, has revolutionized traditional industrial processes, offering a spectrum of opportunities for increased efficiency, productivity, and sustainability.  As shown in Figure \ref{fig:smartmanufacturing}, manufacturing encompasses the seamless convergence of advanced technologies, including the Internet of Things (IoT), artificial intelligence (AI), sensing and control, edge and cloud computing, robotics, among others. Through the strategic integration of these components, manufacturers can optimize operations, streamline production processes, and deliver products of superior quality while responding swiftly to dynamic market demands. 

\begin{figure}
\centering
\includegraphics[width=1\textwidth]{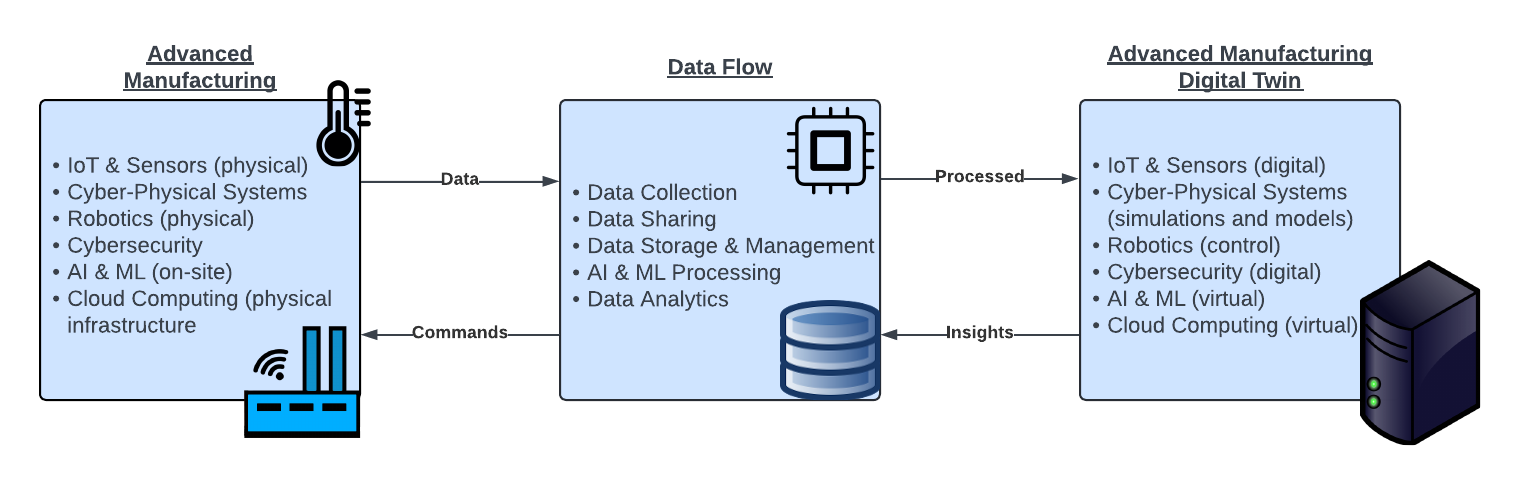}
\caption{Digital Twin for Advanced Manufacturing}
\label{DT_concept}
\end{figure}

Digital twins, virtual replicas of physical assets or systems, have emerged as powerful tools for enhancing operational efficiency, predictive maintenance, and overall decision-making processes \cite{national2023foundational,van2023digital}.  In advanced manufacturing, the concept of the digital twin has gained substantial momentum as a groundbreaking technological advancement, poised to reshape the landscape of manufacturing in both contemporary and future contexts. Acting as a mirror of the real world as shown in Figure \ref{DT_concept}, the digital twin for manufacturing offers a platform for simulating, predicting, and optimizing physical manufacturing systems and processes. By harnessing the power of the digital twin alongside intelligent algorithms, organizations can realize data-driven operational monitoring and optimization, foster innovation in products and services, and diversify their approaches to value creation and business models.

Digital twins offer a broad spectrum of applications across diverse industries and sectors, spanning healthcare, smart cities, manufacturing, supply chain, and so forth. In the realm of healthcare, digital twins are utilized for personalized medicine, the advancement of medical devices, and precision surgical planning. In the domain of smart cities, these digital counterparts contribute to the evolution of intelligent urban endeavors by crafting intricate models of urban infrastructure, transportation grids, and public services. The outcome is astute city planning and judicious allocation of resources.




However, the applications of digital twins, including manufacturing, are likely to confront a critical concern: privacy and security, which are both imperative in this rapidly evolving technological landscape. As the digital counterparts of real-world manufacturing systems become increasingly sophisticated and interconnected, they raise a big number of questions and challenges concerning the protection of sensitive data, the safeguarding of critical infrastructure, and the preservation of individual privacy in manufacturing processes.  To create trustworthy digital twins in manufacturing, ensuring data integrity and cybersecurity is essential. Data integrity preserves the accuracy, consistency, and reliability of information as digital twins are updated and adapted.  Furthermore, achieving trustworthiness in digital twins relies on robust uncertainty quantification methods that account for data variability and enhance decision-making accuracy by providing clear indicators of confidence in simulated outcomes~\cite{thelen2023comprehensive}. Cybersecurity strategies are critical in preventing unauthorized access and ensuring safe model updates, which are especially important in interconnected systems like manufacturing lines. Enhancing trust in digital twins enables them to function as reliable and secure tools for decision-making in advanced manufacturing and resilient supply chain operations.

In addition to addressing the direct cybersecurity (i.e., security and privacy) challenges associated with digital twins in advanced manufacturing, it is equally important to consider the implications of model updates, decision-making processes, and uncertainty quantification within these systems. Ensuring that digital twins can be securely updated and that these updates adhere to privacy standards is crucial for maintaining the integrity and reliability of the manufacturing process. Furthermore, decision-making based on data from digital twins must account for cybersecurity considerations so that people should incorporate robust methods for quantifying uncertainties. This is essential for achieving a high degree of operational efficiency and resilience against potential cyber-threats. Exploring these aspects will further our understanding of how to safeguard the advanced capabilities of digital twins while ensuring their effective deployment in advanced manufacturing.

The rest of the paper is organized as follows. Section 2 addresses the security and privacy associated with data collection for digital twins in advanced manufacturing, emphasizing the imperative for the system to align with the requisites of performance, interoperability, and reliability.  Section 3 addresses the security and privacy issues in data sharing, particularly within the framework of cultivating trust, establishing traceability, upholding data integrity, implementing access control, employing encryption, and delving into the intricacies of centralized versus decentralized data sharing solutions.  Section 4 discusses the potential cybersecurity threats when applying machine learning or deep learning techniques. Cybersecurity issues within model update, decision making, and uncertainty quantification are also discussed. Section 5 addresses system-level security and privacy.  Finally, section 6 demonstrates final remarks and future research goals.

\begin{figure}
\centering
\includegraphics[width=1\textwidth]{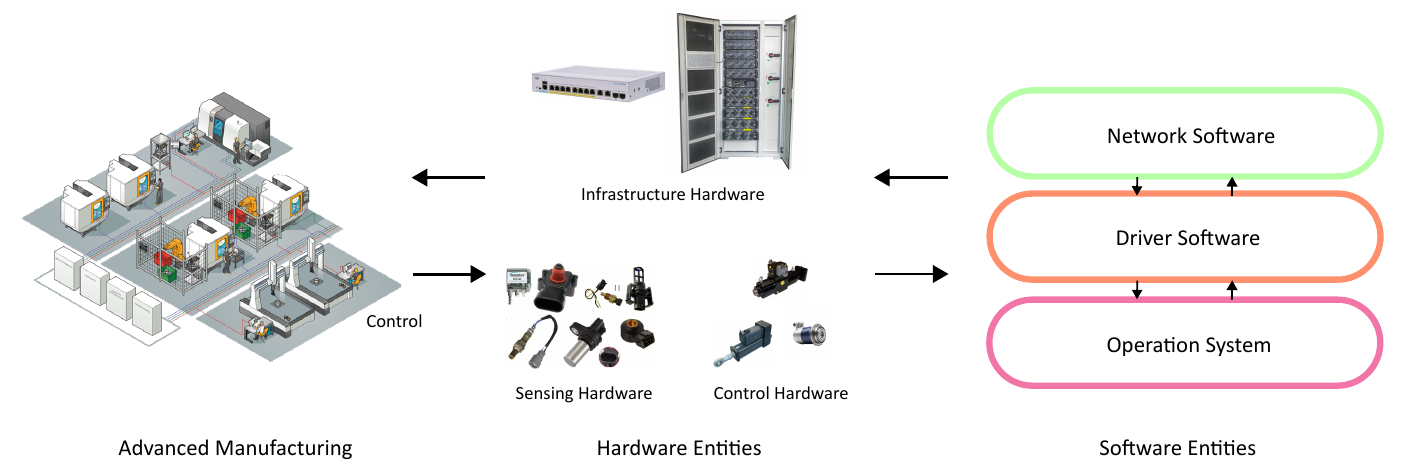}
\caption{Overview of Entities in Data Collection}
\label{Overview of Entities in Data Collection}
\end{figure}

\section{Data Collection}
Manufacturing systems are sophisticated physical entities that are characterized by intricate interconnections among their diverse components. Digital twins enable semi-physical simulation capabilities, effectively curbing the extensive time and costs associated with physical commissioning and reconfiguration. In advanced manufacturing, a physical entity within a factory is linked to the industrial Internet through conventional cyber gateways and is conceptualized as a digital twin in the virtual realm. Each digital twin in cyberspace serves as a representation of its real-world counterpart, mirroring its physical state. The virtual space acts as a repository for streamed data originating from the interconnected physical objects. These data are harnessed to model, simulate, and forecast the status of each tangible entity in the system, especially in the face of dynamic operational conditions.

The life cycle of a digital twin begins in the physical realm, where IoT devices capture dynamic data and formulate control instructions for physical assets. Embedded systems integrate computation, networking, and physical processes, creating a reciprocal influence between computational outcomes and physical events. This integration forms a closed-loop that connects sensors, controllers, and actuators. Controllers synchronously process real-world observations (measurements) and implement command and control instructions to modify real-world behaviors. This initial stage of the digital twin involves both Hardware Entities (HW) and Software Entities (SW). HW resides in the physical domain, while SW acts as a bridge between HW and the data-sharing layer.

As illustrated in Fig. \ref{Overview of Entities in Data Collection}, HW entities in the physical domain fall into three categories: 1) Sensing Hardware, such as IoT sensors, smart meters, and wearable devices, responsible for real-time data collection; 2) Control Hardware, consisting of actuators that execute actions based on feedback from the digital space; 3) Infrastructure Hardware, including grid, networking, and computing infrastructures.


\subsection{Hardware and Software Security}

Digital twin systems face a variety of security challenges at both hardware and software levels. Combining the discussions of hardware and software security emphasizes common challenges such as privilege escalation and insider attacks while summarizing examples of side-channel and privilege escalation threats to avoid redundancy.

\subsubsection{Insider and Privilege Escalation Attacks}
For all hardware devices, insider attacks represent a significant security vulnerability. These devices are often stored in areas accessible to employees, making it easier for insiders to compromise the functionality of sensors, actuators, or power lines. Individuals with full access to Operational Technology (OT) domains can undertake actions that present security threats. They possess the ability to deploy, replicate, or replace IT/OT devices and, in certain instances, maliciously modify software components. Such access enables them to corrupt the physical environment, thereby affecting the digital realm as well. For example, Murillo et al. \cite{10.1145/3442144.3442147} illustrated how manipulating the settings of a Programmable Logic Controller could force hydraulic system water pumps to remain off, leaving a tank empty. This scenario was identified using a digital twin named DHALSIM for detection purposes. Similarly, in \cite{10.1145/3264888.3264892}, researchers altered the logic of a controller to highlight the need for robust online defense strategies.

Privilege escalation attacks often arise when adversaries exploit vulnerabilities in authentication and authorization mechanisms. For example, Triton \cite{xenofontos2021consumer}, malware targeting Triconex controllers, exploited two zero-day vulnerabilities, enabling attackers to elevate privileges, gain memory access, and execute arbitrary code. Attackers with comprehensive access rights to industrial domains may modify configurations, generate false data, or manipulate network traffic \cite{laaki2019prototyping}. These actions can cause significant deviations from standard operations and compromise anomaly detection systems within digital twins \cite{xu2021digital}.

This threat can also originate from the hardware/software supply chain. Unethical manufacturers might introduce compromised components into Cyber-Physical Systems or Industrial IoT devices for various aims, such as creating information leaks, inducing system failures, or undermining asset integrity. These actions affect not only the system's regular operations and its associated digital twins but also potentially damage the organization's reputation.

\subsubsection{Side-channel and Communication-based Attacks}
Side-channel attacks exploit unintended information leakage from computing devices to deduce sensitive data. The concept, pioneered by Kocher \cite{kocher1996timing}, has evolved significantly with the advent of cloud computing. Modern variants, such as cache-timing \cite{ge2018survey} and DRAM row buffer \cite{pessl2016drama} attacks, can be executed remotely in cloud environments. Recently, the spectrum of side-channel attacks has expanded to include acoustics, targeting vibrations of physical entities that inadvertently leak information. For instance, acoustic emissions from keyboards can be triangulated to identify specific keystrokes \cite{pongaliur2008securing}, while automated DNA synthesizers are vulnerable to acoustic side-channel attacks, risking the confidentiality of DNA sequences \cite{faezi2019oligo}.

Hardware-based defenses alone are insufficient to mitigate these risks. Fell et al. \cite{fell2019tad} propose removing conditional branches with distinguishable execution times and replacing primitive instructions with non-deterministic ones to obfuscate timing leakage.

Man-in-the-Middle (MitM) attacks are another major concern in digital twins, particularly in wireless networks that enable synchronization between hardware and digital twins. In \cite{laaki2019prototyping}, researchers exploited mobile networks to cause significant delays in applications like remote surgery control. Insiders may also compromise communication channels by introducing rogue devices, launching routing attacks, or disrupting digital twin traffic \cite{kayan2022cybersecurity}.

\subsubsection{Denial of Service Attacks}
Denial of Service (DoS) attacks target resource depletion in Industrial IoT/Cyber-Physical Systems, hindering automation operations in the physical realm and disrupting simulation processes in the digital realm. These attacks can occur at multiple layers of the TCP/IP stack: jamming at the physical layer \cite{10.1007/978-3-030-37352-8_53}, malware injection at the application layer, or on-the-path attacks at the network layer \cite{ZHANG2019106871}. In advanced manufacturing, where data is transmitted in large quantities, these attacks can result in operational interruptions, compromising the seamless functioning of digital twins \cite{RUBIO2019101561}. Adversaries strategically deplete resources, curtailing automation and simulation activities, and exacerbating vulnerabilities within digital twin systems.

\subsection{Countermeasures}
The interconnection among digital twin models and elements introduces potential security vulnerabilities, often stemming from hardware and software susceptibilities, as detailed in \cite{8822494}. These vulnerabilities may arise from either the absence of robust design practices (security-by-design) or insufficient validation, particularly when incorporating third-party resources. During the development of an advanced manufacturing system, diverse sets of equipment from various manufacturers are often integrated, necessitating a concurrent control design for multiple virtual machines. Traditional control programs, crafted based on engineers' individual experiences, pose challenges for comprehensibility of logical action flow and hinder the swift response to iterative optimization during the control debugging process. The implementation of automated control code generation is paramount, not only facilitating a more straightforward modification and upgrade process but also enhancing collaboration by making the logical flow of actions more accessible to others involved in the development.

Securing the infrastructures integral to the digital twin, including networks, servers, and virtualization systems, is of utmost importance. In this scenario, a comprehensive defense-in-depth strategy forms the fundamental approach for safeguarding digital twin systems. This approach necessitates the integration of robust security mechanisms to safeguard access to digital assets, establishing a layered defense that reinforces the protection of the entire digital twin ecosystem\cite{8972288}.

To establish the first layer of defense, effective measures such as isolation and segmentation can be employed to decouple simulation functions from unauthorized or external access, as suggested in \cite{8822494}. Implementing this involves deploying a range of security elements, including firewalls, proxies, diode communication, secure interconnection devices within virtual networks, adherence to best practices, intrusion detection/prevention systems, and the integration of deception mechanisms. These components collectively form the primary defense mechanisms to fortify the security posture. It's crucial to use predefined access limits and determine the level of trust for every entity engaging with digital twin services. The effectiveness of these mechanisms hinges on their configuration, location, and the responsible entity for their management.

\section{Data Sharing}
After data collection, it is essential in advanced manufacturing that the collected data is shareable without compromising security or privacy considerations. This demands establishing a foundation of trust through secure methods, tamper-proof logs to ensure data fidelity critical aspect of an advanced manufacturing ecosystem, a clear traceability mechanism, access control protocols, and maintaining data integrity to ensure accuracy. The seamless and secure sharing of data across various components of the manufacturing system, such as robots, sensors, and assembly lines, is pivotal for operational efficiency. It is also important to consider the optimal data storage approaches before sharing the data to support the swift and reliable exchange of information required for real-time decision-making in a smart factory. This involves implementing encryption measures to protect the content, employing a cost-effective database infrastructure, and conducting a comprehensive analysis of both centralized and decentralized solutions to determine their merits and decide which one(s) to use.

A solution must not only prevent unauthorized data redistribution but also possess the capability to trace any potential trust breaches. This approach ensures that data remains both accessible and protected throughout its journey, from collection to sharing, a journey that, in advanced manufacturing, is often complex due to the integration of various systems and technologies. Potential threats that need to be considered include unauthorized access by external parties, data interception, insider attacks, manipulation of data, instances of social engineering, and insufficient practices for the  anonymization of data.  The paper \cite{10} also mentions that the data sharing should support high volume data sharing with low-latency in certain scenarios, such as when coordinating across an advanced manufacturing supply chain. In this section, we discuss the requirements for a good data sharing paradigm based on the awareness of security and privacy vulnerabilities, and potential solutions for advanced manufacturing environments, where the stakes are particularly high due to the potentially disruptive consequences of security breaches.  Fig. 4 shows the areas that should be considered when data sharing: 1) how the data is stored, where the data is stored, and data safeguarding, 2) how the data is accessed by authorized persons, and 3) the integrity and provenance of the data.  These will be discussed in the following sections.

\begin{figure}
\centering
\includegraphics[width=0.5\textwidth]{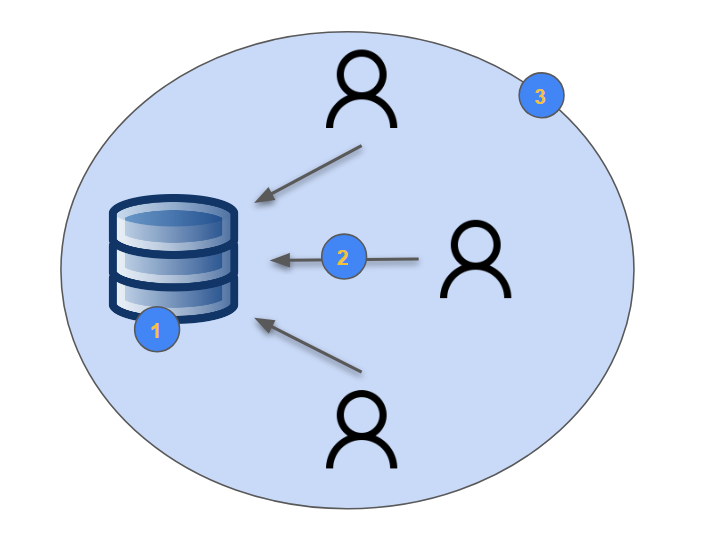}
\caption{Data Sharing Challenges Areas: 1) Data Storage, 2) Data Access, 3) Data Provenance}
\label{system}
\end{figure}

\subsection{Security and Privacy Challenges in Data Sharing}
Data sharing, even in the absence of digital twins, presents an array of security and privacy challenges that are magnified in the context of advanced manufacturing. 

First and foremost, ensuring an accurate historical lineage from the very genesis becomes important, as any modification at any point within the data’s history could compromise both its accuracy and provenance. In an advanced manufacturing setting where decision-making often hinges on the integrity of historical data, this is particularly critical. Document updates have become more frequent \cite{7}, adding complexity to their storage and historical tracking, which is paramount in advanced manufacturing processes for quality control and compliance purposes.

Secondly, the chosen data storage approach—be it a centralized database like Google Drive or a decentralized one incorporating blockchain technology—introduces its distinct set of privacy and security challenges pertinent to the advanced manufacturing milieu.

Thirdly, safeguarding the data necessitates robust techniques like pseudonymization \cite{3} and anonymization \cite{13}, which shield sensitive information from exposure—a critical concern in competitive manufacturing sectors. Federated learning offers significant benefits for maintaining privacy while enabling collaboration. This method preserves data ownership by allowing data owners to contribute model parameters to a central server rather than sharing raw data, significantly mitigating the risk of data breaches. Federated learning, as explored by \cite{zhang2023federated}, enhances privacy and leverages distributed data sources to improve model robustness and accuracy without compromising privacy.

Lastly, it is imperative that only authorized personnel can access and share the stored data. Thus, proper confidentiality, authentication, and access control \cite{8} mechanisms tailored for advanced manufacturing systems are crucial. As digital twins become increasingly integrated into advanced manufacturing, these challenges are further complicated, as additional layers for data leakage emerge during data collection and storage processes.

The rest of this section will discuss various secure data-sharing methods that handle the above-mentioned issues, focusing on their applications in advanced manufacturing. When exploring the types of attacks that can occur during data sharing, \cite{6} discusses potential attacks such as Man-in-the-Middle attacks (MITM), Forgery attacks, and Injection attacks. One proposed solution is the use of blockchain technology.

\subsection{Data Provenance}
When engaging in data sharing, it is important to inspect the provenance of the data, which pertains to the origin and history of the data, ensuring its trail can be followed to ascertain authenticity and integrity. The objective is to establish a robust assurance that the data has not been tampered with and is accurate throughout its journey. Tracking digital twin data from its creation phase and preserving its integrity is critical in advanced manufacturing, ensuring the digital twin accurately reflects its physical counterpart throughout the process.

\cite{1} A blockchain-based framework ensures secure traceability, accessibility, and immutability throughout the digital twin lifecycle, from design to delivery. Smart contracts govern transactions, restricting unauthorized interactions and aligning the twin’s state with its lifecycle phase, such as preventing delivery during testing. This approach enhances synchronization, operational efficiency, and product quality, with potential extensions for continuous improvement cycles.

\cite{5} proposes the concept of a product profile for tracking events throughout a product’s lifecycle, including design, manufacture, usage, and maintenance. Stored on a blockchain, these profiles provide a secure and transparent history of products in advanced manufacturing, enhancing traceability and operational efficiency. Product profiles serve as an additional method for tracking digital twin history, fostering transparency and trust in manufacturing operations.

\subsection{Data Storage}
Data storage constitutes another pivotal dimension of data sharing, particularly in advanced manufacturing where the management of large-scale, complex data is crucial. Within this context, two distinct categories emerge: centralized and decentralized storage mechanisms. While solutions exist for both paradigms, the secure attributes intrinsic to decentralized systems, such as blockchains, position them as the preferred choice for storage, especially for advanced manufacturing applications where security, scalability, and reliability are paramount. Blockchain is known for its decentralized nature as a distributed and immutable ledger database. In the context of a decentralized network that is often the backbone of the advanced manufacturing data architecture, the transmission of data between untrusted digital twins can be authenticated and documented through the blockchain's consensus mechanism. This ensures the security and traceability of the data-sharing process, a critical aspect in intricate and interconnected advanced manufacturing systems. By utilizing smart contracts, digital twins in advanced manufacturing have the ability to collaboratively establish and tailor suitable sharing protocols\cite{6}. This is why most solutions use a decentralized method for storing data before sharing, reflecting the trend in advanced manufacturing towards decentralized data management for enhanced security and efficiency.

The authors of \cite{5} implemented a way of putting all digital twin sensor data along with the product profile onto blockchain because of the hidden security benefits of peer-to-peer, or decentralized, data storage and sharing. Such methods are increasingly relevant in advanced manufacturing where secure, real-time data exchange is vital. Storing data on the blockchain has many benefits, such as cryptography that ensures only eligible participants can access the corresponding data, change-sensitive characteristics of blockchain that ensure data authenticity, and the use of smart contracts to execute certain actions automatically to increase data sharing efficiency, all of which are crucial in advanced manufacturing settings. Peer-to-peer networks are very convenient and quick to share data across eligible parties. Blockchain guarantees data authenticity and verification, providing a robust solution to data management, sharing, and access.

Storing all data on the blockchain might not be practical due to size constraints. A hybrid approach links on-chain metadata to off-chain sensor data using Distributed Hash Tables (DHTs) \cite{11}. Another approach securely records sensor-to-digital twin data transfers in the distributed ledger for backup, with critical updates shared within the blockchain network \cite{12}. \cite{14} proposes a system using cloud computing for efficient data sharing and blockchain to verify and maintain data integrity. The cloud stores the data, blockchain records its hash and logs, and users verify data integrity via the blockchain. Similarly, \cite{17} suggests a model where all data resides in the cloud, and its hash and transactions are stored on the blockchain, though this approach relies on a centralized cloud authority.

In \cite{16}, a model training engine for multi-party data sharing combines blockchain and federated learning. This engine has practical applications in advanced manufacturing and other contexts, enabling secure multi-party data sharing while safeguarding privacy. The encryption scheme proposed is more comprehensive and efficient than conventional mechanisms, making it valuable for handling shared information securely. This study also combines on-chain and off-chain data storage.

\subsection{Safeguarding and Access Control}
Achieving an absolute guarantee of zero data leaks in perpetuity is an impractical aspiration, particularly in the context of advanced manufacturing where the data ecosystem is vast and multifaceted. Therefore, it is prudent to employ safeguarding techniques to fortify data before sharing it with others, a practice that is especially critical in advanced manufacturing due to the sensitive nature of manufacturing data. These techniques include encryption, rigorous access controls, anonymization, pseudonymization, and the integration of blockchain technology. By adopting these safeguards, an additional layer of protection ensures the security and integrity of manufacturing data.

\cite{19} discusses using anonymization as a safeguarding technique  with the advantage that privacy is maintained, which is particularly important in advanced manufacturing where proprietary processes and customer data are involved; however, loss of information and linking attacks are possible. Other privacy-preserving techniques discussed include federated learning and secure multiparty computing (SMPC), both of which are gaining traction in advanced manufacturing due to their effectiveness in maintaining data privacy while enabling collaborative innovation.

Blockchain technology provides significant access control capabilities through smart contracts, which can ensure secure and authorized data access. \cite{11} and \cite{1} demonstrate how smart contracts serve as governing mechanisms for authentication, a crucial requirement in advanced manufacturing involving numerous stakeholders. \cite{2} further explores access control for sharing data with external digital twins or third parties, emphasizing the importance of single-point access control for maintaining tight data security. Standard frameworks like the Extensible Access Control Markup Language (XACML) and tokens such as SAML or OAuth can define and enforce advanced security policies, making them highly relevant for complex manufacturing environments.

Various methods exist for data sharing, each with unique advantages and challenges. One effective method is the publish-subscribe pattern for data sharing, as suggested in \cite{4}. This messaging architecture facilitates timely and efficient communication in software systems. In advanced manufacturing, publishers disseminate messages to a central broker, and subscribers—components of the manufacturing process, including digital twins—express interest in specific topics. The message broker ensures messages reach all relevant subscribers, enabling flexible and scalable data sharing. This decoupling of publishers and subscribers supports responsive, loosely coupled, and event-driven systems, where components can independently react to data changes in a timely manner, critical for advanced manufacturing operations.

\section{Machine Learning and Deep Learning}
Machine learning (ML) and deep learning (DL) are often considered synonymous in recent years, but there are indeed distinctions. ML is a comprehensive artificial intelligence (AI) approach that utilizes diverse algorithms for learning patterns and making predictions.  The ML process typically involves manual feature engineering. In contrast, DL, a subset of ML, utilizes neural networks (NNs) with multiple layers, which automatically learns hierarchical features from raw data, eliminating the necessity for extensive manual feature engineering. DL plays a crucial role in advancing AI across various fields, including computer vision (CV) \cite{voulodimos2018deep}, natural language processing (NLP) \cite{torfi2020natural}, speech recognition \cite{kamath2019deep}, and medical image analysis \cite{esteva2021deep}, which is primarily due to the growing capability of gathering, storing, and processing a vast amount of data \cite{liu2017survey}.

Advanced manufacturing is increasingly integrating ML and DL into the fabric of digital twins, which is evident in several innovative research efforts. Zhang et al. \cite{zhang2020deep} introduce a DL-enabled framework for intelligent process planning in their study. This framework is designed to transform raw materials into final products as specified by product designers through drawings or 3D CAD models. The process is orchestrated within a Digital Twin Manufacturing Cell (DTMC), enhancing both precision and efficiency. Furthermore, Zhou et al. \cite{zhou2020knowledge} develop a comprehensive framework for a Knowledge-Driven Digital Twin Manufacturing Cell (KDTMC). Their approach supports intelligent manufacturing by enabling autonomous operations through sophisticated strategies for perception, simulation, understanding, prediction, optimization, and control. Min et al. \cite{min2019machine} explore the application of ML within the petrochemical industry. They propose a method for constructing digital twins that leverages industrial IoT technologies. This method facilitates a continuous loop of information exchange between the physical factory and its virtual counterpart, optimizing production control.  Finally, Xia et al. \cite{xia2021digital} adopt a data-driven strategy using deep reinforcement learning to promote digital transformation in advanced manufacturing systems. Their approach utilizes digital twins to represent manufacturing cells, which helps in simulating system behaviors, predicting process faults, and adaptively controlling variables to improve system responsiveness and reliability.

To the best of our knowledge, there are {two survey articles} related to digital twins for advanced manufacturing by using ML or DL \cite{sheuly2022machine, rathore2021role}. The primary aim of the survey paper by Sheuly et al. \cite{sheuly2022machine} is to provide a comprehensive review of digital twins in the manufacturing sector, which highlights ML's significant contributions to this field and the incorporation of advanced ML technologies in the design of digital twins, particularly NNs tailored for specific applications. It addresses the lack of systematic literature reviews on various aspects of ML-based digital twins in manufacturing, evaluating this topic from both bibliometric and evolutionary perspectives.  In comparison, the study by Mazhar et al. \cite{rathore2021role} encompasses ML and big data in the context of  digital twins, covering various application domains such as manufacturing, medical, transportation, and energy sectors, while the paper \cite{sheuly2022machine} specifically targets the manufacturing sector and focuses exclusively on ML-based digital twins.

An important contention is that the application of ML or DL models in digital twins for manufacturing will inevitably raise security and privacy concerns, as similarly explored in other studies across domains such as CV or NLP. Generally, {privacy-related concerns} encompass the illicit acquisition or reverse engineering of DL models, the deduction of sensitive training data, and the restoration of identifiable facial images of individuals. Regarding {security}, DL models are susceptible to adversarial attacks that involve slight, imperceptible perturbations introducing the potential for erroneous predictions with significant confidence. This vulnerability represents a critical challenge to investigate when deploying DL models in digital twins for manufacturing.

\begin{figure}[b]
\centering
\includegraphics[width=0.9\textwidth]{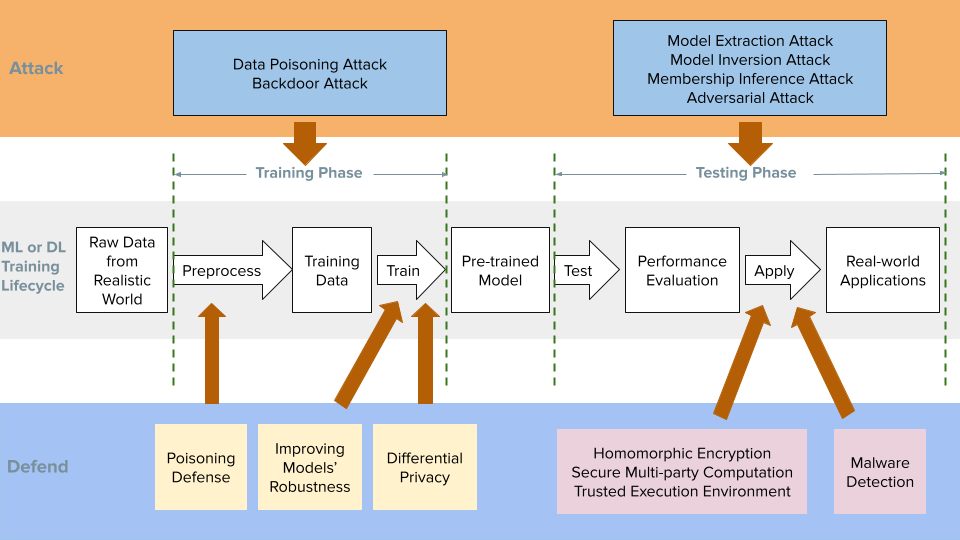}
\caption{The overview of attacks and defenses in ML or DL inspired from the paper \cite{9294026}. We comprehensively review the existing privacy and security issues based on the DL life cycle. In addition, we also analyzed the defense methods. The arrows have the following meanings: Data Poisoning and Backdoor Attacks target at the training phase of the lifecycle, where we could apply different defence techniques (Poisoning Defense, Improving Models' Robustness, and Differential Privacy) to the Preprocessing part or the Training part, respectively. Additionally, Model Extraction, Model Inversion, Membership Inference and Adversarial Attacks target at the testing phase of the lifecycle, where we could apply different defence techniques (Homomorphic Encryption, Secure Multi-party Computation, Trusted Execution Environment and Malware Detection) to the Applying part.}
\label{attacks}
\end{figure}

Therefore, this section gathers relevant literature discussing these concerns, which could be applied to the context of digital twins in advanced manufacturing. Moreover, we introduce cybersecurity in Model Update, and Uncertainty Quantification within digital twins, which are highly related to ML. The remainder of this section will discuss different attacks on privacy and security and how to defend against these attacks  as outlined in Fig. \ref{attacks}.

\subsection{Privacy in ML/DL}
There are usually two privacy threats of DL: 1) model extraction attack; 2) membership inference attack.
Meanwhile, four mainstream defenses are proposed: 1) differential privacy (training phase); 2) homomorphic encryption (training and testing phase); 3) secure multi-party computation (training and testing phase); 4) trusted execution environment (training and testing phase).  See Table \ref{OverviewOfPrivacy} for an overview of privacy threats and their defense types.

\begin{table}[ht]
\centering
\caption{Overview of Privacy Threats and Defense Types.}
\begin{tabular}{|c|c|c|}
\hline
\multirow{2}{*}{\textbf{Privacy Threats}}
& Model Extraction Attack & \cite{tramer2016stealing,wang2018stealing,wang2019attacks,uchida2017embedding,hitaj2018have}  \\
\cline{2-3}
& Membership Inference Attack & \cite{shokri2017membership,long2018understanding,hitaj2017deep,melis2019exploiting,hayes2017logan,nasr2019comprehensive}  \\
\hline
\hline
\multirow{4}{*}{\textbf{Defense Types}}
& Differential Privacy & \cite{xie2018differentially,acs2018differentially,phan2016differential,phan2017preserving,phan2017adaptive,papernot2016semi,papernot2018scalable,triastcyn2018generating}\\
\cline{2-3}
& Homomorphic Encryption & \cite{gilad2016cryptonets,chabanne2017privacy,hesamifard2017cryptodl,bourse2018fast,sanyal2018tapas} \\
\cline{2-3}
& Secure Multi-Party Computation & \cite{aono2017privacy,mohassel2017secureml,liu2017oblivious,rouhani2018deepsecure,juvekar2018gazelle,wagh2018securenn} \\
\cline{2-3}
& Trust Execution Environment & \cite{ohrimenko2016oblivious,hunt2018chiron,hynes2018efficient,chen2018tvm,gu2018securing,hanzlik2021mlcapsule} \\ 
\hline
\end{tabular}
\label{OverviewOfPrivacy}
\end{table}

\subsubsection{Privacy Threats}~\newline
Privacy threats in ML and DL typically manifest through sophisticated attacks aimed at exploiting model vulnerabilities. 

\textbf{Model Extraction Attack:} Introduced by Tramèr et al. \cite{tramer2016stealing}, model extraction attacks seek to replicate the parameters of ML models used in cloud-based ML services. Following this, Wang et al. \cite{wang2018stealing} explored hyperparameter stealing, which targets the theft of hyperparameters critical to the model's performance. Further research by Wang et al. \cite{wang2019attacks} demonstrated that watermarking techniques could inadvertently increase the variance in a model's weight distribution, which could potentially be exploited. Additionally, Hitaj et al. \cite{hitaj2018have} described how adversaries could use stolen models to offer unauthorized Machine Learning as a Service (MLaaS), effectively bypassing detection mechanisms intended to protect legitimate model owners.

\textbf{Membership Inference Attack:} The concept of Membership Inference Attack (MIA) was first put forward by Shokri et al. \cite{shokri2017membership}, highlighting a technique where adversaries train a model to infer whether data samples were used in training another model. This attack is particularly effective against models that have not been adequately generalized. Long et al. \cite{long2018understanding} extended this idea into a Generalized Membership Inference Attack (GMIA) applicable to well-generalized models. Moreover, Hitaj et al. \cite{hitaj2017deep} and Melis et al. \cite{melis2019exploiting} adapted this attack for collaborative DL environments, where they demonstrated that even models developed in a collaborative setting are vulnerable to such privacy breaches. Hayes et al. \cite{hayes2017logan} expanded the scope of MIAs to generative models using Generative Adversarial Networks (GANs) to analyze discrepancies in data distribution. Nasr et al. \cite{nasr2019comprehensive} further refined this approach for a DL setting, offering insights into the robustness of privacy-preserving mechanisms in complex models.

\subsubsection{Mainstream Defences}~\newline
As the landscape of privacy threats in ML/DL has evolved, defensive strategies have been designed to protect against these threats. These defenses are primarily structured around ensuring data privacy and integrity throughout the ML life-cycle.

\textbf{Differential Privacy} \newline
Differential Privacy (DP) has emerged as a cornerstone technique for safeguarding training data against privacy attacks. Abadi et al. \cite{abadi2016deep} pioneered this approach with the introduction of Differentially Private Stochastic Gradient Descent (DPSGD), enabling the training of Deep Neural Networks (DNNs) with enhanced privacy guarantees. Building on this foundation, Xie et al. \cite{xie2018differentially} developed a Differentially Private Generative Adversarial Network (DPGAN), which introduces noise into the discriminator’s gradient to maintain privacy. Similarly, Acs et al. \cite{acs2018differentially} proposed a model combining multiple generative networks, such as Restricted Boltzmann Machines and Variational Autoencoders, to further enhance privacy through a novel method called Differentially Private Generative Model (DPGM).

Expanding the scope of DP, various researchers have focused on objective perturbation techniques. Phan et al. \cite{phan2016differential} introduced the Deep Private Autoencoder (DPA), which perturbs the objective functions to enforce differential privacy. This method was later extended to more complex networks, such as the Private Convolutional Deep Belief Network (PCDBN) \cite{phan2017preserving}, which employs polynomial approximations for non-linear objective functions. Furthermore, Phan et al. \cite{phan2017adaptive} developed the Adaptive Laplace Mechanism (AdLM), optimizing differential privacy for DNNs through adaptive noise generation.

Label perturbation also plays a crucial role in DP. Papernot et al. \cite{papernot2016semi} introduced Private Aggregation of Teacher Ensembles (PATE), which transfers knowledge from multiple teacher models to a student model without compromising data privacy. This technique has been scaled to larger networks and more complex tasks, as demonstrated in subsequent studies \cite{papernot2018scalable, triastcyn2018generating}, which extend PATE's applications to GANs.

\textbf{Homomorphic Encryption}\newline
Another pivotal defense mechanism that allows for computations on encrypted data, thus enabling privacy-preserving model inference. This concept was theoretically demonstrated by Xie \cite{xie2014crypto} and practically implemented in CryptoNets by Gilad-Bachrach et al. \cite{gilad2016cryptonets}, which performed inference on encrypted data using a NN. The methodology has been continuously refined, with enhancements like polynomial approximation of activation functions by Chabanne et al. \cite{chabanne2017privacy} and improvements in computational efficiency in frameworks such as FHE-DiNN \cite{bourse2018fast} and TAPAS \cite{sanyal2018tapas}, which focus on accelerating encrypted computations within NNs. FHE-DiNN leverages the bootstrapping technique introduced by \cite{chillotti2016faster} to achieve strictly linear complexity concerning the depth of the NN.

\textbf{Secure Multi-Party Computation} \newline
Secure Multi-Party Computation (SMC) within DL serves critical functions in scenarios where data privacy is paramount. The first scenario involves data holders who wish to participate in model training or inference without centralizing their sensitive data. Instead, they distribute the data across multiple servers, where each server processes part of the data independently, ensuring that no single server has access to all the information. The second scenario extends this concept to collaborative environments, where multiple stakeholders train a shared model while maintaining the privacy of their individual datasets. Shokri et al. \cite{shokri2015privacy} pioneered this approach by developing a collaborative DL system where participants could learn from each other's data without directly sharing it. They achieved this by training local models individually and then sharing only selected gradients, which minimizes data exposure but raises concerns about potential data leakage from these shared components. Aono et al. \cite{aono2017privacy} addressed this issue by introducing an additive homomorphic encryption technique to safeguard against even the slightest data inference from gradient sharing, though at the cost of increased communication overhead.

Building on these foundations, Mohassel and Zhang \cite{mohassel2017secureml} introduce SecureML, which further refines privacy-preserving techniques in ML. SecureML utilizes a combination of Oblivious Transfer (OT), secret sharing, and Yao's Garbled Circuit (GC) protocol to train NNs securely in a multi-party computational setting. While effective, the original SecureML protocol was limited to simpler NN architectures that did not include convolutional layers. This limitation was overcome by Liu et al. \cite{liu2017oblivious} with the development of MiniONN, a framework that adapts existing NNs for secure computation by transforming them into 'oblivious' networks. This adaptation allows for privacy-preserving predictions with improved efficiency by approximating non-linear activation functions using GC techniques.

Further advancements in SMC were made by Rouhani et al. \cite{rouhani2018deepsecure} through the introduction of DeepSecure, a scalable framework designed to enable privacy-preserving DL inference across distributed systems. DeepSecure leverages both GC and OT protocols to protect client data during computations, effectively balancing security with computational overhead. However, GC's high communication cost remains a challenge. This led to enhancements by Juvekar et al. \cite{juvekar2018gazelle} and Wagh et al. \cite{wagh2018securenn}, who worked on optimizing the balance between homomorphic encryption for linear functions and garbled circuits for non-linear functions. Juvekar et al. improved computational and communication efficiencies by employing a minimal-sized circuit design specifically for non-linear functions, whereas Wagh et al. introduced SecureNN, a more communication-efficient multi-party protocol that dramatically reduces execution times in practical applications.

\textbf{Trusted Execution Environment} \newline
Trusted Execution Environments (TEEs) offer a secure computational space that is isolated from the main processor, specifically designed to protect the execution of sensitive applications. This isolation helps in maintaining the confidentiality and integrity of data and code during execution, making TEEs crucial for addressing privacy and security challenges in DL. Liu et al. \cite{liu2020lightning} highlight the TEE's role in securely managing resource access for trusted applications.

One of the foundational implementations of TEEs in ML is by Ohrimenko et al. \cite{ohrimenko2016oblivious}, who developed a multi-party ML system on untrusted platforms using Software Guard Extensions (SGX) \cite{mckeen2013innovative}. SGX provides robust memory encryption that creates protected areas called enclaves. These enclaves safeguard any data and computations within them from external access, ensuring that even high-privilege processes cannot view or alter their contents. In this system, participants can securely upload and process encrypted models and data, with SGX ensuring that the execution remains confidential and tamper-proof.

Building upon the capabilities of TEEs, Hunt et al. \cite{hunt2018chiron} created Chiron, a system for collaborative ML training under a privacy-preserving framework known as MLaaS (Machine Learning as a Service). Chiron uses multiple SGX enclaves to handle different subsets of data, optimizing training efficiency while ensuring data privacy. However, it limits the interaction with the trained model post-training, which can restrict its applicability for service providers who need broader access to the trained models.

Further extending TEE's utility in DL, Hynes et al. \cite{hynes2018efficient} developed Myelin, a framework that employs the TVM compiler \cite{chen2018tvm} to generate a secure runtime environment for both the training and testing phases of DL models. Myelin utilizes SGX to provide a secure enclave for model training, ensuring that all operations are protected from external threats. Additionally, Gu et al. \cite{gu2018securing} introduced Deepenclave, which cleverly splits DNN processing into two segments: FrontNet and BackNet. FrontNet operates within a secure enclave, while BackNet runs in a less secure environment, thus balancing security with computational efficiency.

The architecture of DNNs and their layer-specific functionalities also benefit from TEEs. Zeiler et al. \cite{zeiler2014visualizing} pointed out that while early DNN layers handle low-level feature detection and can be securely processed within TEEs, the deeper layers that manage more abstract information might not require such stringent security measures. This selective security application helps in optimizing performance without compromising the overall integrity of the system. To further enhance execution efficiency within TEEs, Tramèr et al. \cite{tramer2018slalom} proposed Slalom, a technique that offloads certain computations to untrusted but faster computing devices while ensuring the security of critical operations within the TEE using cryptographic techniques such as Freivalds's algorithm \cite{freivalds1977probabilistic}.

Moreover, for secure offline execution of ML algorithms on client devices, Hanzlik et al. \cite{hanzlik2021mlcapsule} proposed MLCapsule. This innovative approach ensures that data remains local and secure, employing TEEs to execute ML models within a controlled environment, thereby safeguarding the data throughout the processing stages. This encapsulation ensures that even if data must be processed or stored outside secure facilities, the integrity and privacy of the computations are maintained.

\textbf{In summary}, these defenses collectively embody the ongoing innovation and adaptation in the field of protecting ML privacy, reflecting the dynamic interplay between emerging threats and the development of robust protective measures.

\subsection{Security in ML/DL}
With the privacy concerns outlined, it is essential to also consider the security implications of ML and DL.  We will now examine the types of security threats these technologies face and discuss effective strategies to secure ML models against such vulnerabilities.  Table \ref{OverviewOfSecurity} presents a comprehensive categorization of security threats and defense mechanisms pertinent to ML and DL.  It highlights both the adversarial and poisoning landscapes, emphasizing their diversity and the multi-faceted approaches required for robust defense.  In this subsection, we will discuss these different attacks and corresponding defenses.
\begin{table}[ht]
\centering
\caption{Overview of Security Threats and Defense Types.}
\begin{tabular}{|c|c|}
\hline
\multirow{2}{*}{\textbf{Adversarial Attacks}}
& White Box \cite{goodfellow2014explaining,szegedy2013intriguing,kurakin2016adversarial,madry2017towards,papernot2016limitations,carlini2017towards,moosavi2016deepfool,moosavi2017universal,guo2017countering,athalye2018obfuscated} \\
\cline{2-2}
& Black Box \cite{su2019one,athalye2018synthesizing,chen2017zoo,tu2019autozoom,brendel2017decision,brunner2019guessing} \\
\hline
\multirow{3}{*}{\textbf{Adversarial Defences}}
& Pre-processing \cite{wang2017adversary,das2017keeping,guo2017countering,efros2023image,raff2019barrage,liao2018defense,samangouei2018defense,bao2018featurized,mustafa2019image} \\
\cline{2-2}
& Model Robustness Improving \cite{kurakin2016adversarial,chang2018efficient,kannan2018adversarial,liu2018feature,xie2019feature,sun2019adversarial,hosseini2017blocking} \\
\cline{2-2}
& Malware Detection \cite{chen2020stateful,tian2018detecting} \\

\hline
\hline
\multirow{3}{*}{\textbf{Poisoning Attacks}}
& Accuracy Drop Attack \cite{munoz2017towards,yang2017generative,munoz2019poisoning} \\
\cline{2-2}
& Targeted Misclassification Attack \cite{koh2017understanding,shafahi2018poison,zhu2019transferable} \\
\cline{2-2}
& Backdoor Attack \cite{gu2017badnets,liu2018trojaning,chen2017targeted,li2020invisible} \\
\hline
\multirow{3}{*}{\textbf{Poisoning Defences}}
& Against Accuracy Drop Attack \cite{steinhardt2017certified,paudice2018detection} \\
\cline{2-2}
& Against Targeted Misclassification Attack \cite{steinhardt2017certified,paudice2018detection} \\
\cline{2-2}
& Against Backdoor Attack \cite{liu2018fine,chen2018detecting}\\
\hline

\end{tabular}
\label{OverviewOfSecurity}
\end{table}

\subsubsection{Adversarial Attacks}~\newline
\textbf{White-Box Attacks:} 
\begin{itemize}
    \item \textbf{L-BFGS:} The susceptibility of NNs to adversarial examples—crafted by introducing minute perturbations to benign inputs—was first demonstrated by \cite{goodfellow2014explaining}. These perturbations remain imperceptible to the human visual system, yet they can substantially mislead the model's predictions with high certainty.
    \item \textbf{Fast Gradient Sign Method (FGSM):} \cite{szegedy2013intriguing} attributed the existence of adversarial samples to the nonlinearity and overfitting tendencies of NNs. However, \cite{goodfellow2014explaining} illustrated that even basic linear models are vulnerable, and introduced FGSM as an untargeted attack algorithm.
    \item \textbf{Basic Iterative Method/Projected Gradient Descent:} Building upon FGSM, \cite{kurakin2016adversarial} extended FGSM through multiple incremental iterations, creating the Basic Iterative Method (BIM). \cite{madry2017towards} employed Projected Gradient Descent (PGD) iteratively to approximate the p-norm ball around an input and proposed robust adversarial training.
    \item \textbf{Jacobian Based Saliency Map Attack (JSMA):} Focusing on the $L_0$ norm, \cite{papernot2016limitations} introduced JSMA that manipulates a select few pixels within an image to mislead the model.
    \item \textbf{CW Attack:} An optimization-centered adversarial attack known as the CW attack was introduced by \cite{carlini2017towards}, which minimizes the visibility of perturbations by constraining their norms.
    \item \textbf{Deepfool Attack:} \cite{moosavi2016deepfool} developed Deepfool, a classifier-based linearized iterative adversarial technique that generates minimal adversarial perturbations.
    \item \textbf{Universal Adversarial Perturbations:} To affect multiple samples, \cite{moosavi2017universal} proposed the Universal Adversarial Perturbation attack (UAP), identifying a universal perturbation applicable to all training samples.
    \item \textbf{Obfuscated Gradient Attack:} Numerous defenses use obfuscated gradients to hinder attackers \cite{guo2017countering}. However, \cite{athalye2018obfuscated} showed how these defenses can be circumvented, describing three types of obfuscated gradients: shattered, stochastic, and vanish/explode gradients.
\end{itemize}

\textbf{Black-Box Attacks:}
\begin{itemize}
    \item \textbf{One Pixel Attack:} Demonstrating the vulnerability of NNs, the one pixel attack, introduced by \cite{su2019one}, proves that altering just a single pixel can mislead a network model into making incorrect classifications. This highlights the sensitivity of models to minimal input changes.
    
    \item \textbf{Expectation Over Transformation (EOT) Attack:} To counter the loss of adversarial properties under real-world transformations such as rotations and shifts, \cite{athalye2018synthesizing} developed the EOT attack algorithm. This method ensures that adversarial examples maintain their effectiveness even when the input undergoes various transformations \cite{goodfellow2014explaining, guo2017countering, luo2015foveation, lu2017no}.
    
    \item \textbf{Zeroth Order Optimization (ZOO):} Inspired by the CW attack \cite{carlini2017towards}, the ZOO technique by \cite{chen2017zoo} enables black-box adversarial attacks by estimating the gradients based on numerous model output queries, thereby sidestepping the need for direct access tothe internal details of the model.
    
    \item \textbf{Autoencoder-Based Zeroth Order Optimization Method (AutoZOOM):} Enhancing efficiency within black-box settings, \cite{tu2019autozoom} proposed AutoZOOM, a method that uses autoencoder principles to generate adversarial examples more query-efficient.
    
    \item \textbf{Boundary Attack:} Addressing scenarios where obtaining model information is challenging, \cite{brendel2017decision} introduced the boundary attack that relies solely on decision outputs (class labels) rather than detailed model data or confidence scores, emphasizing its applicability in more restrictive real-world scenarios.
    
    \item \textbf{Biased Boundary Attack:} Improving upon the boundary attack, \cite{brunner2019guessing} developed a biased sampling technique that more effectively targets robust models by drawing perturbation candidates from a biased multidimensional distribution, thereby enhancing the efficiency and effectiveness of the attack.
\end{itemize}

\subsubsection{Adversarial Defences}~\newline
Several defenses against adversarial attacks were proposed in recent studies, which grew along three main directions: pre-processing, improving the model robustness, and malware detection. ~\newline

\textbf{Pre-processing Defenses:}
\begin{itemize}
    \item \textbf{Randomization:} Introduced by \cite{wang2017adversary}, this innovative defense strategy randomly neutralizes features within samples. Complementing this, \cite{prakash2018deflecting} utilizes pixel deflection as a defense technique,involving the redistribution of pixel values and incorporates wavelet-based denoising.

    \item \textbf{Image Transformation:} The efficacy of JPEG compression in mitigating adversarial effects was initially explored by \cite{dziugaite2016study}, highlighting its potential in reducing classification errors caused by FGSM \cite{goodfellow2014explaining} generated examples. Further research by \cite{das2017keeping} noted the ability of JPEG compression to eliminate high-frequency signal components, akin to selective blurring. This led to the proposal of a preprocessing module involving JPEG compression to fortify networks against various adversarial attacks. However, \cite{guo2017countering} found that techniques like total variance minimization \cite{rudin1992nonlinear} and image quilting \cite{efros2023image} can outperform traditional methods like JPEG compression \cite{dziugaite2016study}, bit depth reduction \cite{xu2017feature}, and non-local means \cite{buades2005non}. To build a more robust defense, \cite{raff2019barrage} combined several simple defenses, including bit depth reduction \cite{xu2017feature}, JPEG compression \cite{dziugaite2016study}, wavelet denoising \cite{antonini1992image}, mean filtering \cite{huang1979fast}, and non-local mean \cite{buades2005non}, considering the challenge of obfuscated gradients \cite{athalye2018obfuscated}.

    \item \textbf{Denoising Network:} Traditional denoising encoders \cite{vincent2008extracting} are commonly used for removing noise but are less effective against sophisticated adversarial perturbations. Addressing this gap, \cite{liao2018defense} devised the High-level representation Guided Denoiser (HGD) using the U-net architecture \cite{ronneberger2015u} to enhance adversarial defense capabilities.

    \item \textbf{GAN-Based Defense:} Utilizing the principles of Generative Adversarial Networks (GANs) \cite{goodfellow2014generative}, \cite{samangouei2018defense} introduced a defensive framework that trains an adversarial generative network to recreate clean images closely resembling the original while removing adversarial perturbations. Building on this, \cite{bao2018featurized} proposed the Featured Bidirectional Generative Adversarial Network (FBGAN) that focuses on learning latent semantic image features that remain consistent even after perturbations.

    \item \textbf{Image Super-Resolution:} Exploring the potential of image super-resolution, \cite{mustafa2019image} theorized and then demonstrated that a well-trained image super-resolution model could effectively map adversarial samples back onto the manifold of natural images, offering a novel defense mechanism against a range of adversarial attacks.
\end{itemize}

\textbf{Improving Model Robustness:}
\begin{itemize}
    \item \textbf{Adversarial Training:} Initially proposed by \cite{goodfellow2014explaining}, adversarial training seeks to enhance model robustness by exposing the network to adversarial examples during training. \cite{kurakin2016adversarial} further refined this concept by integrating batch normalization \cite{ioffe2015batch}, applying it effectively to the Inception-v3 model \cite{szegedy2016rethinking} and the ImageNet dataset \cite{krizhevsky2012imagenet}. Although successful against single-step attacks, it struggles with iterative attacks \cite{madry2017towards}. To overcome this, \cite{chang2018efficient} developed a dual adversarial samples-based training method effective against both attack types. Additionally, adversarial training strategies by \cite{goodfellow2014explaining}, \cite{huang2015learning}, and \cite{shaham2018understanding} tend to be specialized against certain attack types. In the realm of adversarial training, \cite{madry2017towards} introduced the PGD attack, mainly tested on MNIST \cite{deng2012mnist} and CIFAR-10 \cite{krizhevsky2009learning}. \cite{kannan2018adversarial} extended these methods to ImageNet \cite{krizhevsky2012imagenet}, incorporating output similarity of paired samples into the loss function, demonstrating enhanced robustness and outperforming previous methods like the integrated adversarial training by \cite{tramer2017ensemble}.

    \item \textbf{Regulation:} To enhance feature understanding, \cite{liu2018feature} introduced a model that prioritizes robust features through non-linear attention modules and $L^2$ feature regularization, helping to improve decision-making against minor image alterations.

    \item \textbf{Feature Denoising:} Noting the amplification of adversarial perturbations across network layers, \cite{xie2019feature} advocated for feature denoising to reduce noise accumulation in the network's feature maps.

    \item \textbf{Convolutional Sparse Coding:} Inspired by \cite{heide2015fast} and \cite{choudhury2017consensus}, \cite{sun2019adversarial} introduced a defense that projects adversarial examples into a low-dimensional, quasi-natural image space, aiding in the restoration of natural image properties.

    \item \textbf{Blocking the Transferability:} To counteract the transferability of adversarial samples, \cite{hosseini2017blocking} developed the "empty label" method. This strategy is effective in black-box scenarios, preventing adversarial samples designed for one model from deceiving others with different architectures or trained on dissimilar datasets.
\end{itemize}

\textbf{Malware Detection:}
\begin{itemize}
    \item \textbf{Stateful Detection:} Recognizing the unique nature of cloud-based ML services that primarily function through user queries, \cite{chen2020stateful} introduced a stateful detection mechanism. This innovative black-box defense strategy effectively monitors the state of user interactions to detect anomalies that could indicate malicious activities, thereby enhancing the security of cloud-based ML systems.

    \item \textbf{Image Transformation:} \cite{tian2018detecting} observed that adversarial samples are often more sensitive to transformations such as rotations and shifts compared to natural images, which typically withstand such alterations without loss of integrity. Leveraging this characteristic, they developed a method to detect adversarial examples by applying image transformations and assessing the stability of the output, thus providing a straightforward yet effective tool for identifying adversarial attacks.

    \item \textbf{Adaptive Denoising:} While traditional denoising techniques are effective against significant noise, they can blur images when applied to minor noise, potentially degrading the model's performance. To address this, \cite{liang2018detecting} introduced adaptive denoising methods that include scalar quantization and smoothing spatial filters. These techniques are designed to refine the image processing in the presence of adversarial perturbations, particularly effective in scenarios with low-level noise, thereby preserving the clarity and accuracy of the classification process.
\end{itemize}

Beyond the direct attacks on machine learning models, there is also the risk of poisoning attacks that can corrupt the training process itself. The next section focuses on these attacks, detailing their mechanisms and presenting robust defenses against them.

\subsubsection{Poisoning Attacks}~\newline
Poisoning attacks seriously compromise the integrity of ML models by injecting malicious samples into the training data, thereby reducing model performance or manipulating its predictions. These attacks are particularly dangerous as they subtly alter the learning process and are prevalent in DL. They can be classified into three main categories: accuracy drop attacks, targeted misclassification attacks, and backdoor attacks.

\textbf{Accuracy Drop Attack:}
\cite{munoz2017towards} introduced an innovative poisoning attack using gradient-based optimization that targets a broad range of ML algorithms, including DL architectures. This method calculates gradients through back propagation to modify learning parameters adversely, affecting the entire training process. Building on this, \cite{yang2017generative} developed a strategy using GANs to accelerate the generation of poisoned data. In their GAN framework, an autoencoder acts as the generator to produce corrupted data, while the target model serves as the discriminator, assessing the damage caused by the poisoned data and adjusting accordingly. Furthering this approach, \cite{munoz2019poisoning} created a GAN-based method to produce poisoning samples that closely resemble genuine data points, thus subtly reducing classifier accuracy when included during training.

\textbf{Targeted Misclassification Attack:}
Seeking to manipulate how a model interprets specific inputs, \cite{koh2017understanding} utilized classical influence functions from robust statistics \cite{cook1980characterizations} to examine and influence predictions made by black-box models. Following this, \cite{shafahi2018poison} introduced the clean-label attack, which crafts poisoned samples through feature collisions. Expanding on this technique, \cite{zhu2019transferable} devised transferable clean-label poisoning attacks using a convex polytope approach, enhancing the success rate of attacks within black-box models compared to the feature collision method.

\textbf{Backdoor Attack:}
Given the high costs associated with training models, many opt for outsourcing the training process to cloud servers or using pre-trained models customized for specific tasks. In this context, \cite{gu2017badnets} introduced BadNet, a maliciously trained network that functions normally during training but, when triggered, fails during testing. This model includes backdoor triggers designed with specific pixels and colors, often in simple shapes like squares. \cite{liu2018trojaning} proposed a trojaning attack that modifies an existing model by generating a trojan trigger without needing the original training data. Such attacks, including those by \cite{chen2017targeted} and \cite{li2020invisible}, typically operate under weak threat models where the adversary has limited knowledge about the model and can only insert a few samples, making the triggers nearly invisible to humans while still recognizable to NNs.

\subsubsection{Poisoning Defenses}~\newline
The sophistication and variability of poisoning attacks necessitate robust defense mechanisms that can adapt to evolving threats rather than solely relying on past data. The defense strategies against these attacks are typically categorized based on the type of attack they are designed to counter, such as accuracy drop attacks, targeted misclassification attacks, and backdoor attacks.

\textbf{Defense Against Accuracy Drop/Targeted Misclassification Attack:}
To combat the wide array of potential poisoning attacks, traditional methods based on past attack patterns are often inadequate. Recognizing this, \cite{steinhardt2017certified} developed a proactive defense framework that enhances model resilience by identifying and excluding outlier data points that could potentially skew the model's learning process. This approach aims to safeguard the training set from being compromised. Complementing this, \cite{paudice2018detection} introduced a pre-filtering strategy that employs outlier detection to assess and mitigate the impact of poisoning attacks. This method is particularly effective in diminishing the effects of optimal poisoning attacks, ensuring the integrity of the training data and the robustness of the resulting model.

\textbf{Defense Against Backdoor Attacks:}
Backdoor attacks pose unique challenges as they are designed to activate only under specific conditions, making them difficult to detect using conventional methods. In response, \cite{liu2018fine} pioneered a method known as "fine-pruning." This technique combines pruning— the process of removing redundant or non-critical neurons from the network—to reduce potential vulnerabilities, with fine-tuning, which adjusts the remaining network parameters to maintain performance without the pruned elements. Furthermore, \cite{chen2018detecting} explored the differences in activation patterns between normal and backdoor-triggered inputs within a network. They leveraged these differences to develop Activation Clustering (AC), an innovative approach that detects and isolates corrupted training samples by clustering based on activation patterns, effectively identifying and removing malicious data from the training set.

Having explored the various attacks and defenses in the context of machine learning and deep learning, it is crucial to understand how digital twins can be dynamically updated to respond to these threats effectively, ensuring continuous improvement and adaptation in manufacturing processes.

\subsection{Model Update}
Recent advancements in digital twin technology for manufacturing systems have highlighted various model update methodologies that enhance accuracy and real-time decision-making capabilities via utilizing ML, which might have cybersecurity issues discussed in subsection 4.1 and subsection 4.2. For instance, Tapas Tripura et al. \cite{tripura2023probabilistic} developed a framework for creating and updating digital twins of dynamical systems using sparse Bayesian machine learning.  This method allows for precise updates of digital twin models by incorporating both input and output information from the dynamical system, thereby quantifying uncertainties associated with the updated models. Similarly, M. E. Biancolini and U. Cella \cite{evangelos2019radial} demonstrated the use of Radial Basis Functions (RBF) to update digital models in computer-aided engineering (CAE), effectively capturing actual manufactured geometries and incorporating them into the nominal design.  These updates are crucial for maintaining the integrity and security of the digital twin, ensuring that any discrepancies or anomalies are addressed promptly to prevent potential security risks in manufacturing operations.

Paromita Nath and S. Mahadevan \cite{nath2022probabilistic} proposed a probabilistic digital twin for the laser powder bed fusion (LPBF) process in additive manufacturing.  Their methodology integrates Bayesian calibration to update uncertain parameters and model discrepancies, ensuring that the digital twin remains tailored to the specific part being produced, thus enhancing both security and privacy by adapting to unique manufacturing scenarios.  Jinsong Bao et al. \cite{bao2019modelling} introduced a comprehensive approach for modeling and operations in manufacturing using the Automation Markup Language (AutomationML), which facilitates virtual-physical convergence and enhances production efficiency.

Automated generation and updating of digital twins, as explored by Giovanni Lugaresi and A. Matta \cite{lugaresi2021automated}, leverages data logs to automatically create accurate models of manufacturing systems, streamlining the process and enhancing system performance estimation.  Michael G. Kapteyn et al. \cite{kapteyn2020toward} advanced the field by employing component-based reduced-order models and interpretable ML to update digital twins of unmanned aerial vehicles, allowing for dynamic mission planning in response to structural damage, emphasizing the role of secure updates in operational security.

Henrik Ejersbo et al. \cite{ejersbo2023dynamic} proposed a model-driven approach for the dynamic runtime integration of new models in digital twins, enabling seamless updates without interrupting ongoing operations. Wei Song et al. \cite{song2021real} introduced a real-time digital twin model updating method based on consistency measurement, employing techniques such as Latin Hypercube global searching and greedy local searching to quickly adjust and correct models, enhancing the security posture by rapidly addressing potential vulnerabilities.  
W. Birk et al. \cite{birk2022automatic} reviewed the automatic generation and updating of industrial process digital twins, discussing both machine learning-based and automated equation-based modeling methods. Yiyun Cao et al. \cite{cao2021simulation} described a multi-fidelity framework that uses a low-fidelity neural network metamodel and a high-fidelity simulation model for optimizing digital twins, ensuring real-time decision support while incorporating security layers to protect against malicious attacks and data breaches.

These diverse methodologies, focusing on updating either specific parameters or entire models, significantly enhance the fidelity, utility, security, and privacy of digital twins in manufacturing, thereby improving predictive capabilities, operational efficiency, and secure operations.  With these advanced methodologies in model updating securing the integrity and responsiveness of digital twins, we now turn to explore how these updated models are utilized to drive decision-making processes, further enhancing both the security and efficiency of manufacturing operations.

\subsection{Decision-Making}
The integration of decision-making methodologies into digital twins has shown significant promise in enhancing manufacturing processes, with an added emphasis on applying ML or DL, which might have cybersecurity concerns discussed in subsection 4.1 and subsection 4.2. For example, Lugaresi and Matta \cite{lugaresi2021automated} developed a method to automate the discovery of manufacturing systems and generate digital twins, providing accurate models to estimate system performance efficiently.  Their approach incorporates security protocols to safeguard data integrity and prevent unauthorized system manipulations, which is critical for maintaining the confidentiality and accuracy of the digital twins.  Similarly, Latif et al. \cite{latif2020case} showcased the application of adaptive simulation-based digital twins using reinforcement learning to improve real-time production output and problem-solving on the manufacturing floor.  These systems are designed with robust security features that ensure the protection of sensitive production data and mitigate risks associated with cyber-physical system vulnerabilities.

In the realm of dynamic scheduling, Villalonga et al. \cite{villalonga2021decision} proposed a framework leveraging decentralized decision-making through a fuzzy inference system for cyber-physical production systems.  This method not only optimizes production schedules dynamically but also enhances security by distributing decision-making processes, thus reducing the potential impact of centralized security breaches.  Helman \cite{helman2022digital} highlighted how digital twin-driven approaches can optimize manufacturing procedures by analyzing and simulating real-time production variants, showcasing the flexibility and adaptability of digital twins in modern manufacturing.

Actionable cognitive twins, as described by Rožanec et al. \cite{rovzanec2022actionable}, use knowledge graph modeling to enhance decision-making in production planning, integrating artificial intelligence to provide insightful and actionable decisions, with advanced security measures to protect the underlying data and machine learning models from tampering or theft.  Kunath and Winkler \cite{kunath2018integrating} explored integrating digital twins with decision support systems to improve order management processes, further demonstrating the versatility of digital twins in various manufacturing applications.
Moreover, Kuehn \cite{kuehn2018digital} emphasized the role of digital twins in decision-making for complex production and logistic enterprises, highlighting the potential for multi-criteria decision-making approaches that include security as a key evaluation criterion.  Jeong et al. \cite{jeong2020design} presented a design process for digital twins in production logistics, stressing the importance of data-driven decision-making in optimizing resources and operations, while ensuring that security protocols are adhered to throughout the process.
Advancements in optimization techniques, such as Bayesian optimization, are being incorporated into digital twins to enhance their decision-making capabilities. Zhu and Ji \cite{zhu2023digital} introduced a digital twin-based multi-objective optimization method for the process industries, leveraging Bayesian optimization to achieve better product quality and higher resource utilization, with integrated security measures to protect the optimization algorithms and resultant data.  Nath and Mahadevan \cite{nath2022probabilistic} discussed the use of probabilistic digital twins for additive manufacturing, incorporating Bayesian calibration to manage process variability and uncertainty, while ensuring that security concerns are addressed in the calibration process to maintain model integrity and privacy.

In the context of fault diagnostics, Zhang et al. \cite{zhang2021matrix} employed Bayesian network modeling within digital twins to improve the reliability and accuracy of maintenance services.  This includes robust security features that safeguard the diagnostic processes from external threats.  Ademujimi and Prabhu \cite{ademujimi2022digital} proposed a co-simulation approach for training Bayesian networks, enhancing fault diagnostics in manufacturing systems using digital twins while ensuring that the co-simulation environments are secure against cyber threats.

Cao et al. \cite{cao2021simulation} explored the use of multi-fidelity frameworks in digital twins for simulation optimization, demonstrating the efficiency of Bayesian optimization methods in real-time production planning and scheduling.  These frameworks are designed with enhanced security protocols to ensure that optimization decisions are protected from unauthorized access and manipulation, thus preserving the integrity and confidentiality of the production processes.

As decision-making processes within digital twins become more refined and integrated, the next crucial step involves robustly addressing the uncertainties inherent in these systems.  Uncertainty quantification emerges as an essential tool to help with the reliability and security of decisions made based on digital twin data, ensuring that these decisions are both informed and resilient against potential vulnerabilities.

\subsection{Uncertainty Quantification}
Uncertainty quantification (UQ) plays a pivotal role in enhancing digital twin technologies within the manufacturing sector, addressing both aleatoric and epistemic uncertainties to improve process reliability, accuracy, and decision-making. Aleatoric uncertainty, arising from inherent variability in manufacturing processes, and epistemic uncertainty, stemming from incomplete knowledge, are both critical aspects addressed by UQ methodologies. There is a trend that people utilize modern ML techniques to realize UQ, which might have cybersecurity issues discussed in subsection 4.1 and subsection 4.2

Studies have demonstrated the integration of UQ with dynamic models and adaptive forecasting frameworks to deliver trustworthy predictions in advanced manufacturing environments \cite{murray2023reconceptualizing}.  This integration is crucial for maintaining the integrity and confidentiality of data within digital twins, as it enables the detection and mitigation of potential security threats arising from predictive inaccuracies.  UQ has been utilized to enhance process design and quality control through comprehensive modeling, employing methods such as Bayesian inference, Monte Carlo simulations, Polynomial Chaos Expansion (PCE), and sensitivity analysis \cite{thelen2023comprehensive}.  These techniques provide a robust statistical foundation for securely managing and predicting the behaviors of manufacturing processes under various scenarios.

In the context of Industry 4.0, UQ helps minimize skewed results and ensures robust implementation of optimization and safety enhancement processes using deep learning and surrogate models \cite{rahman2022leveraging}.  This not only improves the reliability of the digital twins but also safeguards against the manipulation of model outputs, thus protecting the systems from security breaches that could compromise operational integrity.

Robust lightweight design in manufacturing benefits from UQ by considering manufacturing quality variations and sustainability requirements, employing robust design methodologies to assess the impact of such variations on fatigue strength \cite{kokkonen2022robust}. Enhanced performance of digital twin systems in manufacturing is achieved through multi-fidelity frameworks that incorporate Polynomial Correlated Function Expansion (PCFE) and Gaussian Process (GP) regression, which help in managing data fidelity \cite{desai2023enhanced, jha2022statistical}. UQ also facilitates real-time monitoring of manufacturing systems by quantifying and characterizing deviation events using statistical analysis and what-if scenarios, ensuring safety and operational integrity \cite{woodcock2021uncertainty}.  This aspect is particularly important for maintaining continuous surveillance against unexpected operational malfunctions or security incidents, enabling timely interventions.  

Edge-cloud digital twin systems benefit from UQ by optimizing server selection for digital twin applications, thus reducing delays and improving convergence rates while ensuring data security \cite{yang2022collaborative}. 
 This optimization is vital for preventing security breaches that could arise from network vulnerabilities or server overloads.  UQ also enhances digital twins for industrial applications, such as elevators, through uncertainty-aware transfer learning, ensuring robust model performance and data integrity \cite{xu2022uncertainty}. Overall, the integration of UQ in digital twin technologies addresses both aleatoric and epistemic uncertainties, fostering more reliable, secure, and efficient manufacturing processes, thus reinforcing the overall resilience and security posture of digital twins in advanced manufacturing settings.

Uncertainty Quantification plays a pivotal role in enhancing the reliability, robustness, security, and privacy of digital twins in advanced manufacturing. Digital twins, when integrated with ML and DL, provide capabilities for real-time monitoring, predictive maintenance, and optimization. However, these systems must effectively manage various uncertainties to ensure their effectiveness, security, and privacy \cite{karkaria2024machine}.  This enhanced focus on security and privacy, especially through the lens of ML and DL, ensures that digital twins not only function efficiently but also safeguard sensitive information and effectively resist malicious activities.

\section{System-level Security and Privacy}
\begin{figure}
\centering
\includegraphics[width=0.6\textwidth]{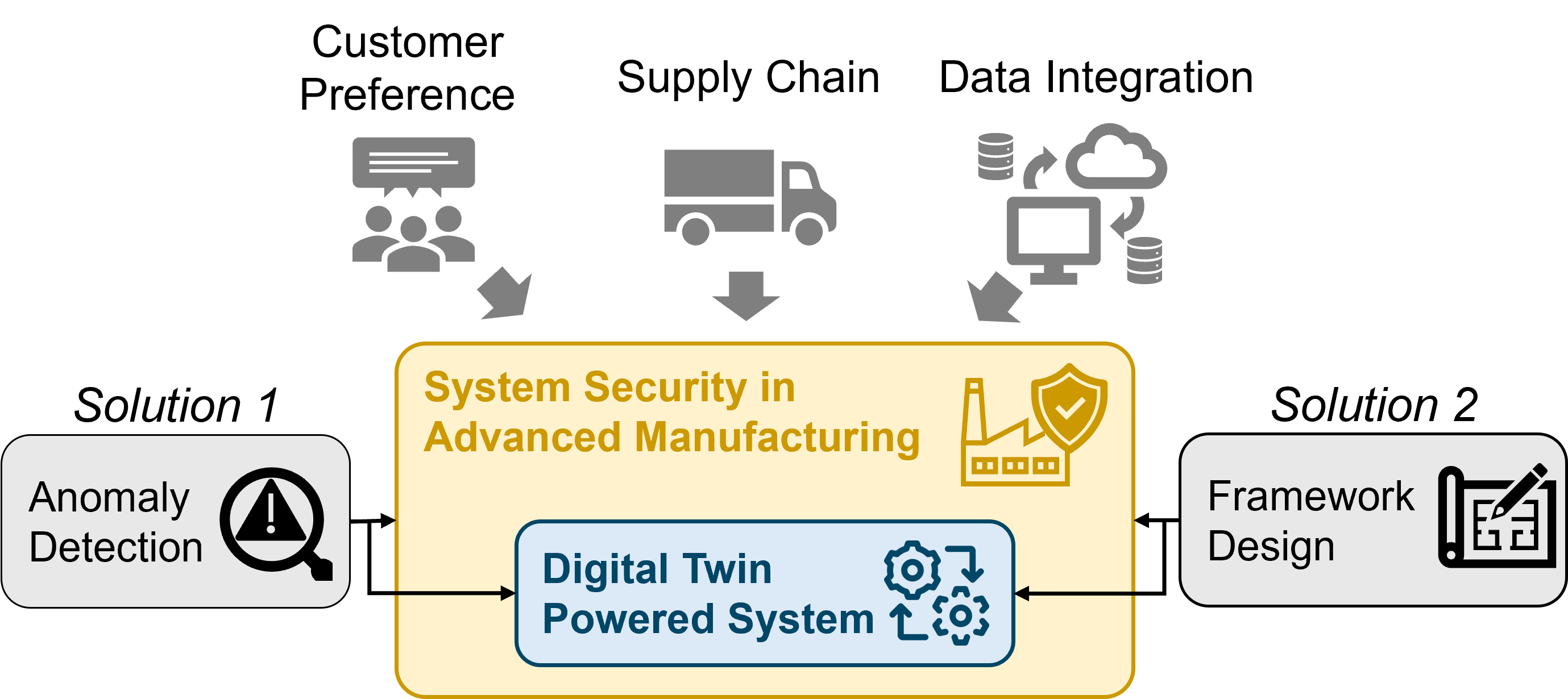}
\caption{Digital Twin System Security in Advanced Manufacturing}
\label{system security in advanced manufacturing}
\end{figure}

An advanced manufacturing system, often synonymous with the Industry 4.0, embodies a fully integrated and autonomous ecosystem tailored to adapt to the dynamic requirements of manufacturing. This system proactively responds to the fluctuating landscapes of production, supply-chain logistics, and customer preferences in real-time \cite{li2022review}. Within this section, as illustrated in Fig. ~\ref{system security in advanced manufacturing}, we delineate two principal methodologies underpinning system security, supplemented by examples. Subsequently, we explore the synergy of these methodologies when integrated with digital twin technology. A comprehensive synopsis of the relevant literature is encapsulated in Table~\ref{table:research-summary}.

\begin{table}[ht]
\centering
\caption{Summary of Research on System Security and Digital Twin in Advanced Manufacturing.}
\begin{tabularx}{\textwidth}{|l|X|l|X|X|}
\hline
\textbf{Study} & \textbf{Type} & \textbf{Digital Twin} & \textbf{Methods} & \textbf{Application} \\
\hline
Sen et al. \cite{sen2023anomaly} & Anomaly Detection & No & Machine Learning, Stream Processing & Industrial Systems \\
\hline
Erba et al. \cite{erba2023assessing} & Anomaly Detection & No & Process-Based Analysis & Resilience against Concealment Attacks \\
\hline
Du et al. \cite{du2023deep} & Anomaly Detection & No & LSTM-Autoencoder, GAN & Industrial Control Systems \\
\hline
Zhao et al. \cite{zhao2023anomaly} & Anomaly Detection & No & PSO-1DCNN-BiLSTM & - \\
\hline
Sun et al. \cite{sun2023industrial} & System Framework & No & Entropy-Based Data Selection, Real-Time Monitoring & Industrial Units \\
\hline
Taibi et al. \cite{taibi2022web} & System Framework & No & Bayesian Gaussian Mixture Models & Network Monitoring \\
\hline
Wilfling et al. \cite{wilfling2022dymola} & System Framework & No & Data-Driven Model, Co-Simulation & - \\
\hline
Makansi et al. \cite{makansi2022condition} & System Framework & No & Supervised Learning, Neural Networks & Industrial Hydraulics \\
\hline
Feng et al. \cite{feng2023model} & Anomaly Detection & Yes & Kalman Filter & Incubator Systems \\
\hline
Zhao et al. \cite{zhao2023dynamic} & Anomaly Detection & Yes & Graph-Based, Attention Mechanism & Complex Systems Monitoring \\
\hline
Li et al. \cite{li2023framework} & System Framework & Yes & Digital Twin Architecture & Forestry \\
\hline
Liu et al. \cite{liu2023digital} & System Framework & Yes & Digital Twin, Multi-Domain Modeling & Industrial Robotics \\
\hline
Fend et al. \cite{fend2022cpsaml} & System Framework & Yes & CPSAML, Code Generation & Mobile CPS Systems \\
\hline
Mehlan et al. \cite{mehlan2022digital} & System Framework & Yes & Virtual Sensor, Data-Driven Approach & Wind Turbine Gearbox Bearings \\
\hline
\end{tabularx}
\label{table:research-summary}
\end{table}

\subsection{System Security in Advanced Manufacturing}
In advanced manufacturing systems, interconnected processes make them vulnerable to cascading failures, or "chain effects," triggered by errors or attacks. Data dependencies and automated decision-making amplify these risks. A multi-layered defense strategy is essential, including network segmentation to isolate critical components, real-time anomaly detection, redundancy for operational continuity, and routine system audits. Robust authentication, personnel training, incident response plans, and AI-based tools further enhance resilience. Two proposed solutions include 1) anomaly detection across digital and physical systems, and 2) an integrated framework for monitoring system status and executing corrective actions.

\subsubsection{Anomaly detection}
In industrial systems, various research efforts have been made to enhance anomaly detection. Sen et al. \cite{sen2023anomaly} introduced an anomaly detection algorithm tailored for large-scale industrial systems, utilizing machine learning and stream processing for real-time data analysis. Erba et al. \cite{erba2023assessing} analyzed the resilience of model-free process-based anomaly detection against concealment attacks, revealing their general susceptibility. Du et al. \cite{du2023deep} proposed an unsupervised detection method using LSTM-Autoencoder and GAN for industrial control systems (ICS), while Zhao et al. \cite{zhao2023anomaly} introduced a PSO-1DCNN-BiLSTM-based solution for ICS.

\subsubsection{Framework design}
On the framework design front, central to digital twin technology, the emphasis is on crafting architectures that mirror both digital and physical system facets, enabling real-time monitoring and action-taking. Sun et al. \cite{sun2023industrial} suggested an entropy-based selection strategy for data modeling and a real-time monitoring framework for industrial units. Taibi et al. \cite{taibi2022web} developed a Bayesian Gaussian Mixture Models-based network monitoring framework, emphasizing passive measurements. Wilfling et al. \cite{wilfling2022dymola} presented a data-driven model generation and co-simulation framework, leveraging Python, Dymola, and the FMI standard. Lastly, Makansi et al. \cite{makansi2022condition} detailed a data-driven condition monitoring approach for industrial hydraulics using supervised learning and neural networks.

\subsection{Digital Twin in Advanced Manufacturing System}
Digital twin technology in advanced manufacturing refers to the creation of a virtual representation or model of a manufacturing process, product, or system, mirroring its real-world counterpart \cite{karkaria2024towards}. This digital replica captures the physical attributes, behaviors, and dynamics of the actual system, allowing manufacturers to monitor, simulate, and analyze operations in real-time. By leveraging data analytics, sensors, and advanced simulation tools, digital twins enable predictive maintenance, process optimization, and enhanced decision-making, bridging the gap between the digital and physical realms of the manufacturing ecosystem. This integration fosters improved efficiency, reduced downtime, and a more agile response to market demands.

\subsubsection{Anomaly Detection for Digital and Physical Systems with Digital Twin}
Anomaly detection plays a pivotal role in maintaining the security and privacy of digital twin systems. It involves the identification of unusual patterns or behaviors that deviate from the norm, which could potentially indicate security threats or system failures. In the context of digital twin systems, anomaly detection can be applied to both the digital representation and the physical system it mirrors. This dual application is particularly relevant in manufacturing settings, where anomaly detection algorithms can monitor the performance of machinery and equipment, identifying any deviations from expected behavior that might indicate a malfunction or security breach.

The paper by Feng et al. \cite{feng2023model} provides an insightful approach to anomaly detection using a Kalman Filter. The authors demonstrate the use of the Kalman Filter through an incubator system and show that it can successfully detect anomalies during monitoring. This approach is particularly effective in real-time monitoring scenarios, where the rapid detection of anomalies can prevent further system damage or data breaches.

Zhao et al. \cite{zhao2023dynamic} introduce a novel approach to anomaly detection in their recent paper. They present a graph-based anomaly detection framework that utilizes an attention mechanism to develop a continuous graph representation of multivariate time series. This is achieved by dynamically inferring edges between time series. This method is especially effective in complex systems featuring numerous interconnected components, such as digital twin systems.

These methods provide a robust foundation for anomaly detection in digital twin systems. However, their application in real-world settings presents several challenges, including the need for large amounts of training data, the complexity of defining what constitutes an anomaly, and the difficulty of integrating these methods into existing systems. Future research in this area should focus on addressing these challenges and developing more efficient and effective anomaly detection methods for digital twin systems.

\subsubsection{Framework Design for Modeling and Monitoring Digital-Physical Systems with Digital Twin}

In their paper, Li et al. \cite{li2023framework} propose a digital twin architecture tailored to a virtual poplar plantation forest system. The framework encompasses both the modeling of the virtual plantation and the analysis of its data. The authors conduct a theoretical analysis of the three primary entities— the physical world, the digital world, and researchers—and explore the mechanisms of their interactions. This research lays foundational groundwork for implementing digital twin technology in forestry.

Focus on industrial robotics, the paper by Liu et al. \cite{liu2023digital} presents a framework that applies a digital twin to industrial robots for real-time monitoring and performance optimization. The framework includes multi-domain modeling, behavioral matching, control optimization, and parameter updating. This approach demonstrates the potential of digital twin technology in enhancing the performance of industrial robots.

Fend et al. \cite{fend2022cpsaml} introduce CPSAML, a language and code generation framework designed for model-driven development of mobile CPS systems. The framework includes a cockpit application that facilitates monitoring and interaction with these systems, showcasing how digital twin technology can be effectively utilized in managing mobile CPS systems.  Mehlan et al. \cite{mehlan2022digital} detail the creation of a virtual sensor designed for online load monitoring and evaluation of the remaining useful life (RUL) of wind turbine gearbox bearings. A digital twin framework integrates data from condition monitoring (CMS), SCADA systems, and a physics-based gearbox model to enable virtual sensing. While these frameworks provide a strong foundation for modeling and monitoring digital-physical systems, real-world implementation faces challenges such as accurately representing systems, ensuring real-time monitoring, and integrating with existing infrastructure. Future research should address these challenges to develop more efficient and effective digital twin frameworks.

\section{Opportunities and Challenges}
The adoption of digital twins in advanced manufacturing offers significant opportunities for enhanced efficiency, optimized production, and improved product quality. By enabling real-time insights, streamlined operations, and predictive maintenance, digital twins drive productivity while facilitating virtual simulations and rapid prototyping to accelerate innovation. Recent advancements, such as dynamic model updates and uncertainty quantification (UQ), further enhance their capabilities. Model updates ensure adaptability to changing conditions, improving accuracy and efficiency, while UQ addresses both inherent and epistemic uncertainties, bolstering the reliability of predictions and decisions. However, these opportunities come with challenges, particularly in security and privacy. The interconnected nature of digital twins and their reliance on sensitive data introduce vulnerabilities to cyberattacks, data breaches, and unauthorized access. Addressing these risks requires robust data governance, secure protocols for transmission and storage, and ethical data handling to safeguard both proprietary and personal information.

Balancing the utilization of digital twins to maximize operational efficiency while safeguarding against potential security breaches and privacy infringements demands a comprehensive understanding of the complex interplay between technology, regulations, and ethical considerations. Addressing these challenges effectively is critical to harnessing the full potential of digital twins in advanced manufacturing and fostering a secure and privacy-preserving environment conducive to sustainable innovation and industrial growth.

Looking ahead, the future of digital twins in advanced manufacturing holds the promise of continued innovation and transformative growth. Anticipated advancements in technology, including the integration of advanced encryption techniques, blockchain solutions, and decentralized data architectures, are expected to fortify the security and privacy infrastructure surrounding digital twins. The implementation of AI-driven security measures and anomaly detection systems will enable proactive threat identification and mitigation, ensuring a resilient defense against emerging cyber threats and unauthorized access. This strategic integration of model updates and decision-making, and uncertainty quantification ensures that digital twins not only optimize manufacturing processes but also adapt and respond to new challenges and opportunities.

\section{Conclusion}
This paper has provided a comprehensive exploration of the intricate landscape of security and privacy in the realm of digital twins within the advanced manufacturing sector. The emergence of digital twins as a pivotal technology in Industry 4.0 has brought forth significant opportunities for advancements in operational efficiency, predictive maintenance, and decision-making processes. However, these benefits are accompanied by considerable challenges, particularly in the domains of data security and privacy.

The interconnected nature of digital twins, along with their reliance on extensive data, introduces potential vulnerabilities that could be exploited by malicious entities. Ensuring robust security mechanisms for data transmission, storage, and access is crucial to prevent unauthorized breaches, cyber-attacks, and to safeguard sensitive information. Additionally, the preservation of privacy is paramount, necessitating stringent data governance, ethical data handling practices, and secure protocols for data sharing and access control.

The future of advanced manufacturing, in the context of digital twins, appears promising with anticipated advancements in technology. The integration of advanced encryption methods, blockchain solutions, and decentralized data architectures is expected to bolster the security and privacy infrastructure. AI-driven security measures and sophisticated anomaly detection systems are set to play a crucial role in proactive threat mitigation, establishing a defense against emerging cyber threats.

As we step into this future, it is imperative to maintain a balanced approach, leveraging the potential of digital twins to maximize operational efficiency while simultaneously safeguarding against security breaches and privacy infringements. The success of digital twins in advanced manufacturing will depend on our ability to navigate this complex interplay of technology, regulations, and ethical considerations, ensuring a secure, privacy-preserving, and sustainable environment for innovation and industrial growth.


\bibliographystyle{ACM-Reference-Format}

\appendix

\end{document}